\documentclass[12pt]{article}
\usepackage{latexsym}

\newcommand{\beq}{\begin{equation}}
\newcommand{\eeq}[1]{\label{#1}\end{equation}}
\newcommand{\bea}{\begin{eqnarray}}
\newcommand{\eea}[1]{\label{#1}\end{eqnarray}}

\begin{document}
\begin{titlepage}
\hfill NYU-TH/01/12/03 hep-th/0112166
\vspace{20pt}

\begin{center}
{\large\bf{HIGGS PHENOMENON FOR 4-D GRAVITY IN ANTI DE SITTER SPACE}}
\end{center}

\vspace{6pt}

\begin{center}
{\large M. Porrati} \vspace{20pt}

{\em Department of Physics, NYU, 4 Washington Pl, New York NY 10003}

\end{center}

\vspace{12pt}

\begin{center}
\textbf{Abstract }
\end{center}
\begin{quotation}\noindent
We show that standard Einstein gravity coupled to a {\em free} 
conformal field theory (CFT) in 
Anti de Sitter space can undergo a Higgs phenomenon whereby the
graviton acquires a nonzero mass (and three extra polarizations). 
We show that the essential ingredients of this mechanism are the discreteness
of the energy spectrum in AdS space, and unusual boundary conditions on the
elementary fields of the CFT. These boundary conditions
can be interpreted as implying the existence of a 3-d defect CFT living
at the boundary of $AdS_4$. 
Our free-field computation sheds light on the essential, model-independent 
features of $AdS_4$ that give rise to massive gravity. 
\end{quotation}
\vfill
\end{titlepage} 
\section{Introduction}
\noindent
Certain compactifications of 5-d gravity to 4 dimensions can be 
interpreted holographically as 4-d gravity coupled to matter. In this 
picture, the Kaluza-Klein excitations
of the fifth dimension are bound states of the matter sector.
Examples of models that admit a  
holographic interpretation are the Randall-Sundrum 
compactifications (RS)~\cite{rs1,rs2} and the Karch-Randall 
compactification (KR)~\cite{kr,kmp2}.
The holographic interpretation of RS was spelled out in several 
papers~\cite{tutti,ahpr,prz}, while that of KR was carried out in~\cite{p2}. 
In~\cite{p2} it was argued that KR is 
dual to a 4-d conformal field theory coupled to gravity 
on $AdS_4$. One of the most surprising aspects of KR is that it
contains massive gravitons {\em only}. 
In the 4-d dual, the mass of the graviton
arises from the graviton self-energy. In the 
holographic approximation, where graviton loops are neglected, this is the 
same as computing the 2-point function of the stress-energy tensor of the CFT.
This computation was carried out in~\cite{p2}.
By giving a purely 4-d interpretation to KR, ref.~\cite{p2}
made clear that KR is the first example of a local 4-d 
field theory in which general covariance does not imply the existence of a 
massless graviton~\footnote{In 3-d a local modification of
the Einstein Lagrangian exists, that makes the graviton massive~\cite{dj}. The 
modification is a Chern-Simons term that does not exist in even dimensions.}. 
The dual of KR is gravity coupled to a 
{\em strongly self-interacting} CFT. Since the graviton mass does not come
from the integrated conformal anomaly~\cite{p2}, it is not clear whether
massive $AdS_4$ gravity is peculiar only to  strongly interacting CFTs.
To answer this question one should compute the 
graviton self-energy in a {\em weakly interacting} CFT. 
In~\cite{p2}, it was suggested to compute the graviton self-energy in the
``least holographic'' model available to us: a free conformal scalar. 

In this paper we perform that calculation and we find
that even when the CFT is free, the graviton can acquire a nonzero mass. 
Even more interestingly, we can pinpoint the property of the CFT that gives 
rise to massive gravity.    

We begin our paper by reviewing in Section 2 
the consequences of general covariance
and Weyl invariance on the two-point function of the stress-energy tensor
(a.k.a. graviton self-energy). 
We show that the Ward identities due to those symmetries only constrain the
self-energy to be transverse and traceless (tt).
We also explain how a graviton mass shows up to quadratic order 
in the tt part of the self-energy. Finally, we 
explain the relation of our findings to the St\"uckelberg formalism. 

In Section 3 we restrict our analysis to free conformal theories in $AdS_4$.
We show why a Higgs-like mechanism that makes the graviton massive can
take place in free field theories in Anti de Sitter space, but not in 
Minkowsky space. Our analysis makes clear why the induced mass of the $AdS_4$
graviton is $O(\Lambda^2/M_{Pl}^2)$.
We also analyze the role of boundary conditions and we find
why graviton mass generation requires non-standard $AdS_4$ boundary 
conditions. Finally, we 
tentatively re-interpret those boundary conditions as due to
the coupling of the 4-d bulk theory to a defect 3-d CFT.

Section 4 contains the explicit calculation of the graviton self-energy when
the matter CFT is a single free, conformally coupled scalar. That calculation
allows us to find the induced graviton mass. 

Section 5 summarizes the findings in this paper, and it 
contains some concluding remarks; among them, a brief discussion of
the model-independent features of our calculation, and comments on possible
generalizations to bigravity models~\cite{kmp2}.

Our metric convention is ``mostly plus,'' $\gamma^0$ is anti-Hermitean, the
$\gamma^i$s are Hermitean.

\section{Ward Identities and the St\"uckelberg Mechanism}
Let us consider a CFT on a 4-d Anti de Sitter space. Let 
$W[g]$ be the generating functional of the correlators of the 
stress-energy tensor. When expanded on an $AdS_4$ background, the one-point 
function $\delta W[g]/\delta g_{\mu\nu}(x)\equiv \langle T^{\mu\nu}\rangle$ 
does not vanish. Indeed, by denoting with
 $E_4=R_{\mu\nu\rho\sigma}R^{\mu\nu\rho\sigma} -4R_{\mu\nu}R^{\mu\nu} + R^2$
the Euler density, and  with 
$C_{\mu\nu\rho\sigma}$ the Weyl tensor, we have
\beq
g_{\mu\nu}{\delta W[g]\over\delta g_{\mu\nu}(x)}=
a C_{\mu\nu\rho\sigma}C^{\mu\nu\rho\sigma} + b E_4 + c \Box R.
\eeq{1}
Here $a,b,c$ are constants that depend on the specific CFT.
On an AdS background $C_{\mu\nu\rho\sigma}= \Box R=0$ and Eq.~(\ref{1}) 
together with the $SO(2,3)$ symmetry of the background implies
\beq
\left.{\delta W[g]\over\delta g_{\mu\nu}(x)}\right|_{g=\bar{g}}= 
{b\over 24}\bar{g}^{\mu\nu}\bar{R}^2.
\eeq{2} 
Background values for $g_{\mu\nu}$, $R$ etc. will be denoted hereafter
by an overbar. In terms of the AdS curvature radius $L$, we have $\bar{R}=
4\Lambda=-12/L^2$.

It is convenient to introduce now another quantity, $\hat{W}[g]=W[g]-(12b/L^4)
\sqrt{g}$, that is stationary on the AdS background:
\beq
\left.{\delta \hat{W}[g]\over\delta g_{\mu\nu}(x)}\right|_{g=\bar{g}}=0.
\eeq{3}
If we couple our CFT to Einstein gravity, the two-point function of the
stress-energy tensor {\em is} the matter contribution to the one-loop 
graviton self-energy, here called $\Sigma$
\beq
 \Sigma^{\mu\nu,\rho\sigma}(x,y)\equiv\left.{\delta^2 \hat{W}[g]\over 
\delta g_{\mu\nu}(x) \delta g_{\rho\sigma}(y)}\right|_{g=\bar{g}}.
\eeq{4}
One may think that the Ward identities due to general diffeomorphisms and Weyl 
invariance would forbid a mass for the graviton, or, at least, relate
it to the conformal anomaly, but this is not the case. To see this, we
expand $\hat{W}[g]$ to quadratic order around the AdS background:
\beq
\hat{W}[g]= \hat{W}[\bar{g}] + {1\over 2} h_{\mu\nu}*\Sigma^{\mu\nu,\rho\sigma}
*h_{\rho\sigma}+ O(h^3),\qquad g_{\mu\nu}=\bar{g}_{\mu\nu} + h_{\mu\nu}. 
\eeq{5}
Here, $A*B\equiv \int d^4x \sqrt{\bar{g}}A(x) B(x)$.
The Ward identities of diffeomorphisms and local conformal (Weyl) invariance
are Eq.~(\ref{1}) and 
\beq
D_{(\mu} \epsilon_{\nu)}*{\delta W[g]\over\delta g_{\mu\nu}}=0.
\eeq{6}
By using the expansion of $\hat{W}[g]$ given in Eq.~(\ref{5}), the two
Ward identities become, to linear order in the metric fluctuations, 
\beq
D_{(\mu} \epsilon_{\nu)}*\Sigma^{\mu\nu,\rho\sigma}*h_{\rho\sigma}=0, \qquad
\bar{g}_{\mu\nu}\Sigma^{\mu\nu,\rho\sigma}*h_{\rho\sigma}=0 .
\eeq{7}
These identities simply state that $\Sigma$ is transverse and traceless. They
do not constrain at all the tt part of the self-energy.

They do not forbid a graviton mass either. Indeed, Ward identities never 
forbid a mass term for a gauge field. 
Let us consider, as a warm-up exercise, the
simpler example of QED in Minkowsky space. In that case, the Ward identities
of QED imply that the photon self-energy is transverse, i.e., 
in momentum space,
\beq
\Sigma_{\mu\nu}(p)= \left(g_{\mu\nu}-{p_\mu p_\nu\over p^2} \right) F(p^2).
\eeq{8}
If $\lim_{p^2\rightarrow 0} F(p^2) \neq 0$ then the photon acquires a nonzero
mass. Equivalently, we can see that mass is allowed by gauge invariance by
using the St\"uckelberg formalism, i.e. by writing the Lagrangian density
of a massive spin-1 field as
\beq
L= {1\over 4} F_{\mu\nu}F^{\mu\nu} + {m^2\over 2}
(\partial_\mu \phi - A_\mu)(\partial^\mu \phi - A^\mu).
\eeq{9}
This Lagrangian density is invariant under the gauge transformation
$\phi\rightarrow \phi + \omega $, $A_\mu \rightarrow A_\mu + \partial_\mu 
\omega$. In the unitary gauge, $\phi=0$, it reduces to the usual Lagrangian
density of a spin-1 field of mass $m$. If we integrate out the 
S\"uckelberg field $\phi$, it reduces instead to a non-local action, that 
contributes to the self-energy $\Sigma$ a term as in Eq.~(\ref{8}), with
$F(p^2)=m^2/2$. To sum up, neither Ward identities nor locality rule 
out a mass term.

In the case of a spin-2 field in $AdS_4$ the basic mechanism at work is the 
same as in our example, even though details differ. 
First of all, the projector over transverse-traceless states is more involved
that in flat space. To find it, it is most convenient to introduce the
Lichnerowicz differential operator $\Delta$~\cite{l} 
On spin-2 fields, it reads
\beq
\Delta h_{\mu\nu}=-\Box h_{\mu\nu} -2R_{\mu\rho\nu\sigma}h^{\rho\sigma} + 
2R_{(\mu}^\rho h_{\nu)\rho}.
\eeq{10}
On the AdS background, $\bar{R}_{\mu\rho\nu\sigma}=(\Lambda/3)(
\bar{g}_{\mu\nu}\bar{g}_{\rho\sigma} -\bar{g}_{\nu\rho}\bar{g}_{\mu\sigma})$, 
$\bar{R}_{\mu\nu}=\Lambda \bar{g}_{\mu\nu}$, and
the Lichnerowicz operator obeys the following properties
\bea
\Delta D_{(\mu} V_{\nu)}&=&D_{(\mu}\Delta V_{\nu)},
\qquad \Delta V_\mu = (-\Box +\Lambda) V_\mu,
\label{11} \\
D^\mu \Delta h_{\mu\nu} &=& \Delta D^\mu h_{\mu\nu},
\label{12} \\ 
\Delta \bar{g}_{\mu\nu}\phi &=& \bar{g}_{\mu\nu}\Delta \phi, \qquad 
\Delta \phi=-\Box \phi,
\label{13}\\
D^\mu \Delta V_\mu &=& \Delta D^\mu V_\mu. 
\eea{14}
These equations state that $\Delta$ commutes with covariant 
derivatives and trace. This is why in its definition 
we could omit the index labeling the degree of the form on which it acts.
Let us consider now the rank-2 symmetric tensor $h_{\mu\nu}$.
Its tt projection must have the form
\bea
h_{\mu\nu}^{tt}&=& \Pi_{\mu\nu}^{\;\;\rho\sigma}*h_{\rho\sigma}\equiv
A(\Delta) h_{\mu\nu} + B(\Delta) 
D_{(\mu}D^\lambda h_{\nu)\lambda} + C(\Delta) D_\mu D_\nu D^\lambda 
D^\rho h_{\lambda\rho} + \nonumber  
\\ && D(\Delta) D_\mu D_\nu h + E(\Delta) \bar{g}_{\mu\nu}h
+ F(\Delta)\bar{g}_{\mu\nu} D^\rho D^\sigma h_{\rho\sigma}.
\eea{15}
Notice that, thanks to the properties of the Lichnerowicz operator given in 
Eqs.~(\ref{11}-\ref{14}) we can treat it as a number, as
it commutes with covariant derivatives and traces.
Using the properties of covariant derivatives on the AdS background, it is easy
to see that tracelessness of $h^{tt}_{\mu\nu}$ implies two equations for the
coefficients $A,..,F$:
\beq
A + 4 E -\Delta D=0,\qquad B-\Delta C
+ 4F=0.
\eeq{16}
Transversality, $D^\mu h^{tt}_{\mu\nu}=0$, implies instead
\beq
2A +(2\Lambda-\Delta) B=0, \qquad
B+ 2(\Lambda-\Delta) C +2 F=0, \qquad
E +(\Lambda-\Delta) D=0.
\eeq{17}
In order to have a projection, $\Pi^2=\Pi$, 
we must normalize $A=1$. The other equations are then solved by 
\bea
&& B={2\over \Delta-2\Lambda}, \qquad C={2\over (\Delta -2\Lambda)
(3\Delta -4\Lambda)}, \qquad D=F=-{1\over 3\Delta -4\Lambda}, \nonumber \\
&& E= {\Lambda -\Delta \over 3\Delta -4\Lambda}.
\eea{18}
As we pointed out, Ward identities allow us to add a term proportional 
to $\Pi$ to the two-point function of the stress-energy tensor
\beq
\Sigma^{\mu\nu\rho\sigma}*h_{\rho\sigma}={c\over 2 L^4}
\Pi^{\mu\nu\rho\sigma}*h_{\rho\sigma}+....
\eeq{19}
Here $c$ is a dimensionless constant. The functional dependence on $L$ is 
fixed simply by dimensional analysis, as the only scale appearing in $\Sigma$
is the AdS curvature $L$.
As in the spin-1 example given earlier, a nonzero $c$ signals that the graviton
acquires a nonzero mass. To see this, we couple the CFT to dynamical AdS 
gravity. Denote with $(16\pi G)^{-1}K$ ($G$=Newton's constant)
the bare graviton kinetic term and integrate out the CFT. 
Assume that graviton loops can be neglected. The 
dressed kinetic term then becomes $(16\pi G)^{-1}K+\Sigma$, 
and the linearized equation of motion of the graviton is
\beq
[(16\pi G)^{-1}
K_{\mu\nu}^{\;\;\rho\sigma} + \Sigma_{\mu\nu}^{\;\;\rho\sigma}]*h_{\rho\sigma}
=0.
\eeq{20}
On tt fields $(Kh)^{tt}_{\mu\nu}=- (m^2/2)h^{tt}_{\mu\nu}$, and the 
equation of motion reduces to
\beq 
[-(16\pi G)^{-1} m^2 +c/L^4]h^{tt}_{\mu\nu}=0, 
\eeq{20a}
thus giving the value $16\pi G c/L^4$ for the graviton square mass.
Notice that if $c$ is $O(1)$, 
the order of magnitude of the graviton mass is as in~\cite{kr}.   

Notice also that the expression for $\Pi$ contains the
non-local term $B$, that can be interpreted as the propagator of a spin-1 
Goldstone boson. Indeed, $B$ has a pole whenever a transverse vector $A_\mu$
exists such that $(\Delta -2\Lambda)A_\mu=0$. In other words, the term
proportional to $B$ in Eq.~(\ref{15}) plays for
spin 2 on AdS the same role of the term $p^\mu p^\nu/p^2$ in Eq.~(\ref{8}). 
In variance with Minkowsky space expectations, our Goldstone vector is 
{\em not} massless. In fact, a transverse, massless spin 1 in $AdS_4$ obeys
the equation $\Delta A_\mu=0$, instead of  
$(\Delta -2\Lambda)A_\mu=0$. The latter equation states that the Goldstone 
vector is in a massive $SO(2,3)$ representation. We will identify that 
representation in the next Section; this one concludes by showing that
the projection $\Pi$ can be obtained from a local Lagrangian by introducing
a St\"uckelberg vector field, in exact parallel with the case of spin 1 in 
flat space.   
The St\"uckelberg vector $A_\mu$
is introduced into the linearized action of a massive
spin-2 field (the Pauli-Fierz Lagrangian~\cite{pf}) by replacing everywhere
the spin-2 field $h_{\mu\nu}$ with $h_{\mu\nu} + D_{(\mu}A_{\nu)}$.
The Pauli-Fierz action on an $AdS_4$ background is then~\cite{p} 
\beq
S=S_L[h_{\mu\nu}] + \int d^4x \sqrt{-\bar{g}}{c\over 4 L^4} 
(h_{\mu\nu}^2 -h^2).
\eeq{21}
Here $S_L[h_{\mu\nu}]$ is the Einstein action with cosmological constant,
\beq
S_E[g_{\mu\nu}]={1\over 16 \pi G} \int d^4x 
\sqrt{-g}[R(g) - 2\Lambda], \qquad g_{\mu\nu} = 
\bar{g}_{\mu\nu} + h_{\mu\nu}
\eeq{22} 
linearized around the Einstein-space background $\bar{g}_{\mu\nu}$. 
The substitution $h_{\mu\nu}\rightarrow h_{\mu\nu} + D_{(\mu}A_{\nu)}$
maps $S$ into
\beq
S_L[h_{\mu\nu}] + \int d^4x \sqrt{-\bar{g}}{c\over 4 L^4}[
(h_{\mu\nu} + D_{(\mu}A_{\nu)})^2 -(h+2D_\mu A^\mu)^2].
\eeq{23}
By integrating out the vector field $A_\mu$, this action reduces to 
$S_L[h_{\mu\nu}]+ (c/ 4 L^4) h*\Pi*h$.

The action in Eq.~(\ref{23}) is invariant under the 
linearized diffeomorphisms
\beq
h_{\mu\nu}\rightarrow h_{\mu\nu} + D_{(\mu} \epsilon_{\nu)} , \qquad
A_\mu \rightarrow A_\mu -\epsilon_\mu.
\eeq{23'} 
 
It is far form evident that the St\"uckelberg mechanism can be made
fully covariant; here, we limited ourselves to its linearization around a
fixed background. The holographic interpretation of the KR model shows, on
the other hand, that
a complete covariantization of the  St\"uckelberg mechanism is in fact 
possible.
The same conclusion would follows if we were 
able to give a mass to the graviton by
coupling Einstein gravity to a free CFT. The very possibility of this
occurrence is due to some peculiar properties of $AdS_4$ that we analyze next.

\section{$SO(2,3)$ Representations and the Gravitational \\ Higgs Mechanism
in $AdS_4$.}
In the previous Section we showed that diffeomorphism invariance does not
forbid a mass term for the graviton. One cheap way to introduce such a mass 
would be to add by hand a term proportional to $\Pi^{\mu\nu\rho\sigma}$ into
the graviton self-energy. This is the same as adding to the stress-energy 
tensor of matter a term $D_{(\mu}A_{\nu)}$, which is 
conserved if $A_\mu$ obeys the equation of motion $(\Delta -2\Lambda)A^\mu=0$.
By thus changing the theory, we modify by hand its infrared behavior, by
adding three extra degrees of freedom. Nothing guarantees us that this change
make sense beyond the linearized level. 
Full consistency on the other hand {\em is} guaranteed if we find the extra
degrees of freedom needed to give mass to the graviton in the stress-energy
tensor of matter. In the case of a free field theory, where $T_{\mu\nu}$ is 
quadratic, this means finding the Goldstone vector as a bound state in 
the product of two free fields. In Minkowsky space this is obviously absurd
since non-interacting two-particle states form a continuum. In Anti de Sitter 
space, instead, free particles do form bound states, since the AdS energy
is quantized. To proceed further we need to review some facts about 
positive-energy representations of the AdS isometry group, $SO(2,3)$.

These representations were classified in~\cite{e} (see also~\cite{n} for a 
clear review). In the decomposition $SO(2,3)\rightarrow SO(2)\times SO(3)$, 
the generator of $SO(2)$ is the $AdS_4$ energy, while angular momentum is 
given by the generators of $SO(3)$. A unitary, irreducible, positive-weight 
representation of $SO(2,3)$ (UIR), $D(E,s)$, is labeled by the energy $E/L$ 
and spin $s$ of its (unique) lowest-energy state. 
$E$ is the energy measured in units of the $AdS$ curvature radius $L$. 
Free fields form irreducible representations of $SO(2,3)$. A
conformal scalar can belong to either the $D(1,0)$ or the $D(2,0)$. 
A conformal (massless) spin-1/2 fermion belongs to a $D(3/2,1/2)^{\pm}$,
while a massless vector (also conformal) belongs to a 
$D(2,1)^{\pm}$~\cite{f,bf}.
The label $\pm$ denotes the two possible parities of the UIR.

Massless representations of spin $s> 0$ 
have $E=s+1$~\cite{f,bf,n}. Massive unitary 
representations of spin larger than zero have $E>s+1$. In the limit
$E\rightarrow s+1$, the UIR $D(E,s)$, $s\geq 1$ 
becomes reducible~\cite{e,f}:
\beq
D(E,s) \rightarrow D(s+1,s) \oplus D(s+2,s-1), \qquad E\rightarrow s+1.
\eeq{24}
Eq.~(\ref{24}) encodes the group theoretical aspect of the Higgs phenomenon in
$AdS_4$: when a spin-$s$ field, $s\geq 1$, becomes massive, it ``eats''
a spin-$(s-1)$ boson. Notice than for $s=0$ this boson is in a $D(3,0)$, i.e. 
it is a minimally-coupled scalar~\cite{bf}. For spin 2, it is a {\em massive}
vector in the $D(4,1)$~\footnote{Recall that in Minkowsky space, instead, the
Higgs phenomenon for a spin 2 requires a massless vector and a massless 
scalar. They together provide 3 degrees of freedom, as it does our
massive vector in $AdS_4$.}. 
Notice that the wave equation obeyed by a vector $A_\mu$ in the $D(4,1)$ is
exactly what we found in the previous Section: $(\Delta -2\Lambda)A_\mu =0$.

The next question we have to address is whether a $D(4,1)$ appears in the
stress-energy tensor of a free CFT. Since the stress-energy tensor of a free
field theory is quadratic in the fields, the $D(4,1)$ can only be in
$T_{\mu\nu}$ if it appears in the tensor 
product of the UIRs to which the fields
belong. Let us examine separately free conformally-coupled fields of spin 1,
1/2, and 0.
\begin{description}
\item{spin 1}
Massless spin-1 fields are conformal; they belong to the $D(2,1)$~\cite{f,bf}.
The tensor product of $SO(2,3)$ UIRs was found by Heidenreich 
in~\cite{h}. For $D(2,1)$, he found
\bea
D(2,1)\otimes D'(2,1)&=& \sum_{n=0}^\infty D(4+n,0) \oplus 
\sum_{n=0}^\infty D(4+n,1) \oplus  \nonumber \\
&& \sum_{S=0}^\infty \bigg[D(4+S,2+S)
\oplus \sum_{n=0}^\infty 2D(5+S+n,2+S)\bigg].
\eea{25}
In our case, since we are tensoring two identical bosons, some of the
representations that appear in the tensor product above are absent.
For instance, the ground state of the
$D(4,1)$ that appears in the tensor product above is antisymmetric in its 
arguments ($\sim \epsilon_{ijk}A^iA^j$), so it is forbidden by Bose 
statistics. This means that the entire $D(4,1)$ is absent.
\item{spin 1/2}
The massless (conformal) spin-1/2 field belongs to the $D(3/2,1/2)$.
Tensoring two different $D(3/2,1/2)$, Heidenreich finds~\cite{h}
\bea
D(3/2,1/2)\otimes D'(3/2,1/2) &=&  \sum_{n=0}^\infty D(3+n,0) \oplus
\sum_{S=0}^\infty \bigg[D(3+S,1+S)\oplus \nonumber \\ &&
\sum_{n=0}^\infty 2D(4+S+n,1+S)\bigg].
\eea{26}
In this tensor product, the $D(4,1)$ appears twice. By taking into account
Fermi statistics when tensoring two identical representations, we get rid of 
one of them. The other one cannot appear in the stress-energy tensor since it 
has the wrong parity. To arrive at this result we first notice that the 
stress-energy tensor of a free CFT made of 
several fields of spin $s\leq 1$ is given by the sum 
$T_{\mu\nu}= \sum_iT^i_{\mu\nu}$.  The $i$-component of this sum is the
stress-energy tensor of either a real vector, a real scalar, or a Majorana
fermion. To preserve the Majorana condition ($\psi=C\psi^*$, $C=$ charge 
conjugation), the field $\psi(x)$ must transforms as follows under parity: 
$\psi(t,\bf{x})\rightarrow \eta \gamma^0 \psi(t,-\bf{x})$, $\eta=\pm 1 $.
The fermion field $\psi$ can be expanded in spherical 
waves. Its positive-frequency part (with respect to the 
global AdS time $t$) is~\cite{bf}
\beq
\psi_{pf} = \sum_{k=0}^\infty\sum_{n=0}^\infty e^{-i\omega t/L}
a_{\omega j m}
\chi_{\omega j m}^\pm,\qquad j=1/2+k,\qquad \omega=1 +j+n. 
\eeq{27}
The operators $a^\dagger_{\omega j m}, a_{\omega j m}$ 
respectively create and annihilate states  
of definite energy $\omega/L$ and angular momentum $j$, belonging to the
$D(3/2,1/2)^{\pm}$.
The superscript $\pm$ labels the parity of the representation. 
More precisely,  when $\psi$ belongs to the $D(3/2,1/2)^+$, the 
spherical waves $\chi_{\omega j m}^+$ in Eq.~(\ref{27}) 
transforms under parity as
$\chi_{\omega j m}^+\rightarrow i(-)^{\omega-3/2}\chi_{\omega j m}^+$. 
Analogously, for $D(3/2,1/2)^-$,   
$\chi_{\omega j m}^-\rightarrow -i(-)^{\omega-3/2}\chi_{\omega j m}^-$.

The parity a UIR 
is fixed by the parity of its ground state. The assignments given
above show immediately that the parity of the ground state of the
$D(4,1)$ in the tensor product of either $D(3/2,1/2)^+ \otimes D(3/2,1/2)^+$ 
or $D(3/2,1/2)^- \otimes D(3/2,1/2)^-$ is +1. This is the parity of a 
{\em pseudo-vector}, while the $D(4,1)$ contained in $T_{\mu\nu}$ must be a 
true vector, with parity $-1$.  This can be seen most easily by noticing that
$T_{\mu\nu}$ is a true tensor and that 
the $D(4,1)$ we are after must appear in it as follows
\beq
T_{\mu\nu}=D_{(\mu}A_{\nu)} +.... ,\qquad (\Delta -2\Lambda)A_\mu=0.
\eeq{28}
Equivalently, we may notice that with a single Majorana fermion we cannot 
form a vector, as $\bar{\psi}\gamma^\mu\psi=0$, but we can form the 
pseudo-vector $\bar{\psi}\gamma^\mu\gamma^5\psi$.

\item{spin 0}
Scalars belong to $D(E,0)$, $E\geq 1/2$.
The tensor product of two spin zero representations of $SO(2,3)$ is~\cite{h}
\beq
D(E_1,0)\otimes D(E_2,0)= \sum_{S=0}^\infty \sum_{n=0}^\infty
D(E_1+E_2+S+2n,S).
\eeq{29}
Here $E_1,E_2>1/2$. When $E=1/2$, the representation degenerates, becoming a 
singleton~\cite{ff,d}, namely a representation that propagates only boundary
degrees of freedom and cannot be represented as a standard 
local field living in the bulk of $AdS_4$. We will not consider it further.
When $E_1=E_2=E>1/2$, a $D(4,1)$ exists in the tensor product 
$D(E,0)\otimes D'(E,0)$ only for $E=3/2$. If the two representations are
identical, $D(4,1)$ is eliminated by Bose statistics [its would be ground 
state is in reality a descendant belonging to the $D(3,0)$].
\end{description}
This is not the end of the story, since the $D(4,1)$ appears in the tensor
product of two representations of different energy. In particular, as it is
evident from Eq.~(\ref{29}), it appears in the product $D(1,0)\otimes D(2,0)$.
As we mentioned earlier, a conformal scalar belongs to either $D(1,0)$ or 
$D(2,0)$. We can obtain a $D(4,1)$ by taking two different 
conformal scalars, $A,B$, one belonging to $D(1,0)$, the other to $D(2,0)$.
Equivalently, we can form a vector out of the two scalars: 
$A_\mu\sim A D_\mu B$. This is not what we need though, as the stress-energy
tensor does not mix the field $A$ with $B$. What we need is more 
unconventional.

When we stated that a conformal scalar belongs to either a $D(1,0)$ or a 
$D(2,0)$ we implicitly assumed certain conditions at the boundary of $AdS_4$.
These boundary conditions are spelled out~\cite{bf}; they amount to ask 
that no momentum or energy escape through the boundary. We will refer to 
them as reflecting boundary conditions. Recent studies of of 
the KR model~\cite{kr,kr2,p2,dwfo,br} have made clear that these boundary 
conditions are not the
most general physically meaningful ones. For instance, the holographic
interpretation of the KR model {\em demands} that energy and momentum  
freely pass through the $AdS_4$ boundary into a ``mirror'' $AdS_4$ 
space~\cite{br}. Equivalently, energy and momentum are absorbed and released
into the 4-d bulk by a 3-d CFT living on the boundary of $AdS_4$. 
For us, this means that we can relax the boundary conditions of~\cite{bf},
i.e. that we can allow the conformal scalar to belong to the reducible 
representation $D(1,0)\oplus D(2,0)$. By doing this, the stress-energy tensor
does contain fields in the tensor product $D(1,0)\otimes D(2,0)$, and it may
even contain a $D(4,1)$. 

Whether the $D(4,1)$ is really in $T_{\mu\nu}$ can only be found by an explicit
calculation of the self-energy. This is the subject of the next Section.
\section{Through a Glass, Clearly: KR Boundary Conditions and the Graviton
Self-Energy}
\subsection{KR Boundary Conditions}
We call KR 
the boundary conditions that allow a conformal scalar to freely
pass through the boundary of $AdS_4$ into a mirror space obtained as follows:
consider the static Einstein Universe, which is topologically $S_3\times R$.
By a Weyl rescaling $g_{\mu\nu} \rightarrow \Omega^2 g_{\mu\nu}$, 
the Einstein universe is mapped into two adjacent $AdS_4$
spaces, joined at their common boundary. 
If $\phi$ solves the equations of 
motion of a conformal scalar in the Einstein-Universe background, then 
$\Omega^{-1}\phi$ solves the equations of motion of the scalar in $AdS_4$.
A complete set of solutions in the Einstein Universe gives 
{\em two} complete sets of solutions in $AdS_4$, namely, $D(1,0)$ and $D(2,0)$.
If we want to describe a scalar free to move from one AdS into the other, 
as required for instance for the holographic 
interpretation of the KR model, we must keep {\em both} sets of modes.
More general boundary conditions can be imposed if we put a 3-d defect CFT at
the boundary of $AdS_4$.  
\subsection{The Scalar Propagator}
Consider $R^5$ with pseudo-Euclidean metric $\eta=diag(-1,-1,+1,+1,+1)$.
Anti de Sitter space is the covering space of the hyperboloid
$X^M X^N \eta_{MN}=-L^2$, $M,N=0,..,4$.
By $SO(2,3)$ invariance, the scalar propagator $\Delta_E(X,Y)$ 
is a function of $Z\equiv X^M Y_M/L^2$ only. 
$\Delta_E(Z)$ obeys the equation
\beq
[(1-Z^2) \partial_Z^2 -3Z\partial_Z + E(E-3)]\Delta_E(Z)=0;
\eeq{30}
The mass of the scalar is related to $E$ by the equation $L^2 m^2=E(E-3)$.
The solution of Equation (\ref{30}) that vanishes at large $Z$ is a 
hypergeometric
\beq
\Delta_E(Z)= rZ^{-E}F(E,E-1;2E-2;1/Z).
\eeq{30a}
The normalization constant $r$ is fixed by requiring that at 
$Z\rightarrow 1$ $\Delta_E(Z)$ reduces to the properly normalized flat-space
Green's function. As shown in~\cite{aj}, this condition gives
\beq
r={1\over 4\pi^2 L^2}{\Gamma(E)\Gamma(E-1)\over \Gamma(2E-2)}.
\eeq{30b}
For conformal coupling, $E=1$ or $2$, and the normalized 
solution of Eq.~(\ref{30}) reduces to
\beq
\Delta(Z)= {1\over 4\pi^2L^2}
\left(\alpha {1\over Z^2-1} + \beta {Z\over Z^2-1}\right). 
\eeq{31}
When the scalar is in the $D(1,0)$, $\alpha=0, \beta=1$; when it is in the
$D(2,0)$, $\alpha=1,\beta=0$. KR boundary condition give instead 
$\alpha=\beta=1/2$. This can be seen as follows. The scalar propagator
in any space is 
\beq
\Delta(x,y)=\sum_{i} {1\over \lambda_i} \phi_i (x) \phi_i(y), \qquad
(-\Box+m^2) \phi_i(x)=\lambda_i \phi_i(x).
\eeq{32}
Call $\psi_i$ the eigenmodes of Eq.~(\ref{32}) that form a
$D(1,0)$, and  $\chi_i$ the eigenmodes that form the $D(2,0)$. 
They can be thought as the modes of $S_3\times R$ restricted to a half 
3-sphere. A complete set of normalized modes on $S_3$ is 
$\{ 2^{-1/2}\psi_i, 2^{-1/2}\chi_i\}$~\footnote{The factor $2^{-1/2}$ ensures 
that the modes on $S_3$ are normalized to one when 
the modes on the half-sphere 
have unit norm, since $\psi_i,\chi_i$ are either symmetric or antisymmetric
under the reflection that maps one hemisphere into the other.}, so
the propagator on $S_3\times R$ is 
\bea
\Delta(x,y)&=&{1\over 2}\sum_{i} {1\over \lambda_i} \psi_i (x) \psi_i(y)+
{1\over 2}\sum_{i} {1\over \lambda'_i} \chi_i (x) \chi_i(y), \nonumber \\
(\Box+2/L^2) \psi_i(x)&=&-\lambda_i \psi_i(x),
\qquad (\Box+2/L^2) \chi_i(x)=-\lambda'_i \chi_i(x).
\eea{33}
When restricted to $AdS_4$, Eq.~(\ref{33}) reduces to Eq.~(\ref{31}) with 
$\alpha=\beta=1/2$. 
We will keep $\alpha,\beta$ generic in most of our calculations.
\subsection{Tensor Fields in Homogeneous Coordinates} 
We could compute the graviton self-energy in intrinsic coordinates, 
using the techniques of ref.~\cite{hfmmr,aj}, but it is much simpler
to use the embedding of $AdS_4$ in $R^5$ given in the previous Subsection, and
to promote all 4-d fields into 5-d homogeneous fields. That technique was
developed in ref.~\cite{f}. 
Consider first a spin-2 field of mass $L^2m^2\equiv E(E-3)$,
represented by a symmetric tensor $h^{\mu\nu}(x)$. 
The embedding of $AdS_4$ into $R^5$ defines 5 coordinates
$X^M(x)$ obeying $X^M(x)X_M(x)=-L^2$. The 5-d tensor field $h^{MN}(X)$ is
then defined as the homogeneous field of degree $N$ that on the hyperboloid 
reduces to
\beq
h^{MN}(x)=\partial_\mu X^M(x) \partial_\nu X^N(x) h^{\mu\nu}(x).
\eeq{34}
By construction it obeys
\beq
X^N\partial_N h^{AB}(X)=N h^{AB}(X), \qquad X^M h_{MN}(X)=0. 
\eeq{35}
The 5-d indexes, $M.N$ etc. are raised and lowered with the flat metric 
$\eta_{MN}$, the degree of homogeneity $N$ is arbitrary.

As shown in~\cite{f}, the 4-d equations of motion are equivalent to the 
following 5-d equations ($\partial^2=\partial_M \partial^M$):
\beq
[X^2\partial^2-(N+E)(N-E+3)]h_{MN}=0,\qquad 
\partial_M h^{MN}=0,\qquad h^M_M=0 .
\eeq{36}
Equivalently, the space of fields that solves Eqs.~(\ref{36}) is $D(E,2)$ for
$E>3$.
For $E=3$, $h_{MN}= X^2\partial_{(M} A_{N)} + (2-N) X_{(M} A_{N)}$ also 
solves Eqs.~(\ref{36}) when $A_M$ obeys 
($A\cdot B \equiv A^MB_M$)
\beq
[X^2 \partial^2 - (X\cdot \partial)^2 -3 X\cdot \partial +4]A_M=0,\qquad
X\cdot A =0, \qquad   \partial \cdot A=0, \qquad X\cdot \partial A_M=(N-1)A_M.
\eeq{37}
$A_M$ is the gauge mode, generating the $D(4,1)$. This follows from the very
definition of gauge mode. More generally, recall that a spin-1 field 
belongs to the $D(E,1)$ ($E\geq 2$). One can then show~\cite{f} that 
its 5-d equations of motion are
\beq
[X^2 \partial^2 - (X\cdot \partial)^2 -3 X\cdot \partial +E(E-3)]V_M=0,
\qquad X\cdot V =0, \qquad   \partial \cdot V=0.
\eeq{38}
These equations show once more that the gauge mode belongs to the $D(4,1)$.

Eqs.~(\ref{37}) allow us to find the operator $P$
that projects the symmetric
tensor $h_{MN}$ over tt modes, i.e. the 5-d equivalent of the operator 
$\Pi^{\mu\nu\rho\sigma}$ given in Eqs.~(\ref{15},\ref{18}). The projector $P$
decomposes as $P=I + P_1 +\sum_i P_0^i$. $I$ is the identity operator on
symmetric tensors, $P_1$ is the projector over spin-1 states, and the $P^i_0$ 
are projectors over spin-0 states. We are interested in finding $P_1$, as it
is the term that we will need to detect the presence of 
a Goldstone vector in the graviton self-energy. 
To find $P_1$, we can compute the scalar product of $P$ in between symmetric 
tensors that are not only $X$-transverse, $X^Mh_{MN}=0$, but also
traceless and double divergenceless: 
\beq
h^M_M=\partial_M\partial_N h^{MN}=0.
\eeq{38'}
On these fields the projector simplifies considerably.
Explicitly, 
\bea
P_{AB}^{CD}h_{CD}&=& h_{AB} -2[X^2\partial_{(A}\eta_{B)} +(2-N)X_{(A}\eta_{B)}]
+ ... \label{39}, \\
\eta_A &=& [X^2 \partial^2 - (X\cdot \partial)^2 -3 X\cdot \partial +4]^{-1}(
\partial^L h_{LA} +...).
\eea{40}
The ellipsis denote terms that vanish in the scalar product. To define the
scalar product $h*P*h$
we have to promote the 4-d integration oven $AdS_4$ to 5-d.
This is done by promoting 4-d scalars $S(x)$ to homogeneous
5-d scalars of arbitrary degree $N$
\beq
X\cdot \partial \hat{S}(X)= N \hat{S}(X), \qquad \left. \hat{S}(X) 
\right|_{X^2=-L^2}=S(x).
\eeq{41}
The 4-d integration is then extended by using the identity 
\beq
\int d^4x \sqrt{\bar{g}}S(x)=  \int d^5X \delta(X^2+L^2) \hat{S}(X)\equiv
\int d\mu(X) \hat{S}(X).
\eeq{42}
The scalar product $h*P*h$ on fields obeying Eq.~(\ref{38'}) finally reads
\beq
h*P*h= \int d\mu \{h^{AB}h_{AB} +2 X^2\partial_C h^{CA}[X^2\partial^2 - 
(X\cdot \partial)^2 -3 X\cdot \partial +4]^{-1}\partial^D h_{DA}\}.
\eeq{43}
This equation identifies the projector over the spin-1 state: it is the term
proportional to $(\partial\cdot h)^2$.

We need one last property before we embark in the computation of the graviton
self-energy, namely the matrix element 
\beq
\langle X| [X^2\partial^2 - 
(X\cdot \partial)^2 -3 X\cdot \partial +4]^{-1}|Y\rangle.
\eeq{44} 
When the two points $X,Y$ lie on the hyperboloid $Y^2=X^2=-L^2$, 
the matrix element is a function of $Z\equiv X\cdot Y/L^2$ only. It obeys
Eq.~(\ref{30}) with $E=4$ so it is equal to $-\Delta_4(Z)$. For 
$Z\rightarrow \infty$ Eq.~(\ref{30b}) gives
\beq
\Delta_4(Z)= {1\over 40\pi^2 L^2}
Z^{-4} +O(Z^{-6}), \qquad Z\rightarrow \infty.
\eeq{45}
\subsection{The Graviton Self-Energy}    
We finally come to the heart of this paper: the computation of the graviton 
self-energy due to a free conformally coupled scalar. Since 
we neglect graviton loops in our computation, the
only contribution to the self-energy comes from the 2-point correlator of
the matter stress-energy tensor
\beq
h*\Sigma*h= \int d^4 x \sqrt{\bar{g}(x)}\int d^4 y \sqrt{\bar{g}(y)}
h^{\mu\nu}(x)\langle T_{\mu\nu}(x) T_{\rho\sigma}(y)\rangle h^{\rho\sigma}(y).
\eeq{46}
A free conformal scalar on $AdS_4$ obeys the equation of motion 
$(\Box -2\Lambda/3)\phi=0$. Its stress-energy tensor is
\beq
T_{\mu\nu}= \partial_\mu \phi \partial_\nu \phi -
{1\over 2}\bar{g}_{\mu\nu}\partial_\lambda\phi D^\lambda\phi - 
{1\over 6} [D_\mu D_\nu - \bar{g}_{\mu\nu}(\Box-\Lambda)]\phi^2.
\eeq{47} 
We can simplify the calculation of the self-energy by evaluating 
Eq.~(\ref{46}) on traceless, double divergenceless tensors 
\beq
h^\mu_\mu=D_\mu D_\nu h^{\mu\nu} =0.
\eeq{48}
On these configurations, Eq.~(\ref{46}) becomes
\beq
h*\Sigma*h= \int d^4 x \sqrt{\bar{g}(x)}\int d^4 y \sqrt{\bar{g}(y)}
h^{\mu\nu}(x)\langle \partial_\mu \phi(x) \partial_\nu\phi (x)
\partial_\rho \phi (y)\partial_\sigma\phi(y) \rangle h^{\rho\sigma}(y).
\eeq{49}
Since $\phi$ is a free field, we use Wick's theorem and Eq.~(\ref{31})
to find
\beq
h*\Sigma*h= 2\int d^4 x \sqrt{\bar{g}(x)}\int d^4 y \sqrt{\bar{g}(y)}
h^{\mu\nu}(x) h^{\rho\sigma}(y) {\partial\over \partial x^\mu} 
{\partial\over \partial y^\rho} \Delta(Z)
 {\partial\over \partial x^\nu} 
{\partial\over \partial y^\sigma} \Delta(Z)  .
\eeq{50}
Now we convert Eq.~(\ref{50}) into a 5-d equation using the results of the 
previous Subsection. Namely, we define a 5-d field $h_{MN}$ by Eq.~(\ref{34})
and we transform the 4-d integration into a 5-d one using Eq.~(\ref{42}).
Recalling the definition of the coordinate $Z=X\cdot Y/L^2$, and setting
$L=1$ henceforth, we arrive at
\bea
h*\Sigma*h&=&2\int d\mu(X)\int d\mu(Y) h^{AB}(X)h^{CD}(Y) [\Delta'(Z) \eta^{AC}
+Y^AX^C\Delta''(Z)][\Delta'(Z) \eta^{BD} \nonumber \\
&& +Y^BX^D\Delta''(Z)], \qquad '\equiv {d\over dZ}.
\eea{51}
As previously explained, we want to see if Eq.~(\ref{51}) contains 
a term proportional to $\partial \cdot h \Delta_4(Z) \partial \cdot h$. To
see that, we first write Eq.~(\ref{51}) as a sum of three pieces
\bea
h*\Sigma*h&=&2\int d\mu(X)\int d\mu(Y)(A+B+C), \label{52} \\
A&=& h^{AB}(X)h_{AB}(X) [\Delta'(Z)]^2, \label{53} \\
B&=&h_{AB}(X) h^{BD}Y^A{\partial \over \partial Y^D} 
[\Delta'(Z)]^2, \label{54}\\
C&=& h^{AB}(X) h_{CD}(Y)X^CX^D {\partial Z \over \partial X^A} \Delta''(Z)
{\partial\over \partial X^B}\Delta'(Z).
\eea{55}
By integrating by part repeatedly the functions $B$ and $C$, 
we cast Eq.~(\ref{51}) into the desired form 
\bea
h*\Sigma*h &=& 2\int d\mu(X)\int d\mu(Y) {\partial \over \partial X^A}h^{AB}(X)
{\partial \over \partial Y^C}h^C_B(Y) [F(Z)-4G(Z)]+..., \label{56} \\
F'(Z)&=& [\Delta'(Z)]^2, \qquad G'''(z)=[\Delta''(Z)]^2.
\eea{57}
In this equation, we omitted all terms not proportional to 
$(\partial_A h^{AB})(X) (\partial^C h_{CB})(Y)$.

Now we can find if the stress-energy tensor of 
our theory contains a vector in the $D(4,1)$. We just need to find if
$F-4G$ contains a term proportional to the propagator of the vector.
That propagator decays as $Z^{-4}$ at large $Z$ [see Eq.~(\ref{46})], so we
need to find if such a term is contained in $F-4G$. The constants of 
integration in the definition of $F$ and $G$ must be chosen so that they decay
at large $Z$. If we recall the definition of $\Delta(Z)$ given in 
Eq.~(\ref{31}) and we expand this function
in powers of $1/Z$ we find that there is only one term proportional to 
$Z^{-4}$:
\beq
F-4K= -{1\over 5 (4\pi^2)^2} \alpha\beta Z^{-4} + ....
\eeq{58}
Notice that this term vanishes when the b.c. are purely $E=1$ or $E=2$, as it
should.
Eq.~(\ref{58}) shows that mixed boundary conditions, 
with both $\alpha$ and $\beta$ nonzero, do indeed give rise to a 
Goldstone vector. This happens in particular with the KR conditions, 
$\alpha=\beta=1/2$. Upon coupling to gravity, this vector supplies the 
extra polarizations needed to make the graviton massive. To find the graviton
mass induced by this Higgs mechanism, we first recall that 
the large-$Z$ behavior of the propagator $\Delta_4(Z)$ is given by 
Eq.~(\ref{45}), thus
\beq
F-4K= -{1\over 2\pi^2} \alpha\beta \Delta_4(Z) + ....
\eeq{60}
Finally, we re-introduce $L$ in our formulas,
recall Eqs.~(\ref{19},\ref{20}), and find: 
\beq
m^2= 16\pi G {2\alpha\beta\over \pi^2 L^4}.
\eeq{61} 
As predicted by dimensional analysis in Section 2, the graviton mass
is proportional to $L^{-4}$ i.e. to $\Lambda^2$.
We emphasize again that the functional dependence on $L$ is model-independent,
as it follows only from two simple assumptions: 1) graviton loops can be 
neglected in the computation of the 
graviton self-energy; 2) matter is conformal. The first 
assumption guarantees that the self-energy is independent of $G$; 
the second guarantees that $L$ is the only mass scale of the matter theory.
Clearly, the numerical coefficient in Eq.~(\ref{61}) {\em is} model 
dependent already in the free theory, since it depends on the boundary 
conditions! Even when the boundary conditions are chosen to be KR 
($\alpha=\beta=1/2$), no non-renormalization theorem is known to us that 
protects Eq.~(\ref{61}). It is not surprising, therefore, 
that the proportionality
constant in Eq.~(\ref{61}) does not coincide with what was found in the KR 
model~\cite{kr,p2}. After all, KR is dual to a {\em strongly interacting}
CFT, while here we examined a {\em free} CFT.
\section{Conclusions}
In this paper we re-examined the possibility of giving a mass to the graviton
in Anti de Sitter space. 

We pointed out that Ward identities do not forbid a 
graviton mass, at least in linearized gravity. We then proceeded to examine
the conditions that allow a gravitational Higgs mechanism in AdS, 
and we gave a model-independent estimate of the graviton mass induced by a
CFT.

Next, we recalled  
that unlike Minkowsky space, $AdS_4$ allows even a free theory to form bound 
states, owing to the discreteness of the AdS energy spectrum. We investigated
free CFTs with spin not greater than 1. We found that, in order
to have a Goldstone particle in the stress-energy tensor, we had to impose
non-standard (i.e. non-reflecting) boundary conditions on the fields of our
free CFT. Similar boundary conditions are not only allowed, 
but indeed necessary, to interpret holographically the KR 
model~\cite{kr,p2,kr2,br}. 

We considered next a free conformal scalar in $AdS_4$, and we found 
that, even in that very simple example, nonstandard boundary conditions did 
produce a Goldstone boson that gives a mass to the graviton, when the CFT is 
coupled to gravity. 

The present calculation found that a graviton mass is generated by coupling 
standard gravity to a free CFT. In~\cite{kr,p2}, it was shown that the same 
phenomenon happens when gravity is coupled to a {\em strongly} interacting CFT.
In both cases the key ingredient is the boundary conditions imposed on the CFT.
They must allow energy and momentum to flow in and out of $AdS_4$ through its
boundary. This phenomenon can be interpreted as due to a 3-d defect conformal 
field theory located at the boundary of $AdS_4$. It is curious that two
vastly different cases --free CFT versus strongly-interacting CFT-- in which
the graviton becomes massive have in common just the choice of 
boundary conditions. We may speculate that this
fact point out to the possibility of finding a model-independent setting
for the gravitational Higgs effect on AdS, that only depends on a choice of
boundary dynamics for the field theory of matter.

Finally, we may ask if the ``bigravity'' model of ref.~\cite{kmp2} can also be 
explained purely in terms of 4-d physics with nonstandard boundary conditions.
We recall that in~\cite{kmp2} a region of $AdS_5$ was bounded by {\em two} 
$AdS_4$
branes. In that model one finds, besides the usual massless graviton, 
a second spin-2 field that couples as the graviton and has mass 
$O(\Lambda^2/M_{pl}^2)$. When one of the two branes is sent to infinity, the 
coupling of the massless graviton vanishes while that of the massive 
graviton remains finite. In that limit, the model becomes KR. 
Clearly, the spectrum of bigravity cannot be reproduced by {\em any} choice of
the parameters $\alpha,\beta$ in Eq.~(\ref{31}), as long as they are constant.
Nevertheless we suggest that a more complicated choice of boundary conditions,
where $\alpha,\beta$ become functions of the $AdS_4$ energy, may give rise to
bigravity. This choice of boundary conditions can be thought of 
as describing the dynamics of the 3-d defect CFT living at the boundary of 
$AdS_4$, that must partially reflect and partially absorb bulk fields.
    
\vskip .2in
{\bf Acknowledgments}\vskip .1in
\noindent
We would like to thank N.N. Khuri for useful conversations. 
This work is supported in part by NSF grant PHY-0070787.

\end{document}